\documentclass[journal]{IEEEtran}
\usepackage{cite}
\usepackage{amsmath,amssymb,amsfonts}
\usepackage{algorithmic}
\usepackage{graphicx}
\usepackage{textcomp}
\usepackage[binary-units]{siunitx}
\usepackage{xcolor}
\usepackage{array}
\newcolumntype{P}[1]{>{\centering\arraybackslash}p{#1}}
\usepackage{url}
\DeclareUnicodeCharacter{2212}{-}

\begin{document}

\title{3D Homogenized T-A Formulation for Modeling HTS Coils}

\author{Carlos~Roberto~Vargas-Llanos
        and~Francesco~Grilli
\thanks{C.R. Vargas-Llanos and F. Grilli are with the Institute for Technical Physics (ITEP) of the Karlsruhe Institute of Technology (KIT), 76131 Karlsruhe, Germany (e-mail: francesco.grilli@kit.edu).}
}

\markboth{Journal of \LaTeX\ Class Files,~Vol.~XX, No.~X, Month~20XX}%
{Shell \MakeLowercase{\textit{et al.}}: Bare Demo of IEEEtran.cls for IEEE Journals}

\maketitle

\begin{abstract}
The estimation of losses in high-temperature superconductors (HTS) is fundamental during the design of superconducting devices since losses can strongly influence the cooling system requirements and operating temperature. Typically, 2D finite element analysis is used to calculate AC losses in HTS, due to the lack of analytical solutions that can accurately represent complex operating conditions such as AC transport current and AC external applied magnetic field in coils. These 2D models are usually a representation of an infinitely long arrangement. Therefore, they cannot be used to analyze end effects and complex 3D configurations. In this publication, we use the T-A homogenization in 3D for the analysis of superconducting coils. This allows simulating complex geometries such as racetrack coils. We show that this approach has lower computation time than the currently available 3D homogenization of the H formulation. The approach is first validated against measurements and 2D axisymmetric solutions. Then, it is used to estimate losses and study the electromagnetic behavior of a racetrack coil.
\end{abstract}


\IEEEpeerreviewmaketitle

\section{Introduction}
\IEEEPARstart{T}{he} electrical properties of high-temperature superconductors (HTS) have inspired several applications in different fields such as electrical machines \cite{snitchler_10_2011},\cite{dolisy_fabrication_2017}, \cite{frauenhofer_basic_2008}; fault current limiters \cite{okakwu_application_2018}; magnets for scientific research \cite{berrospe-juarez_estimation_2018};  energy storage \cite{rong_developmental_2017} and transmission \cite{fietz_high-current_2016}. The design of these devices usually requires an electromagnetic analysis that allows establishing rated characteristics as well as studying the behavior under different operating conditions. Moreover, losses in the superconducting tapes and wires must be estimated to design the cooling system. These AC losses can be decisive for the practical and economic realization of superconducting devices.

Several analytical solutions have been developed to estimate losses in HTS tapes \cite{brandt_type-ii-superconductor_1993}, \cite{norris_calculation_1970} and infinite stacks of tapes \cite{mawatari_critical_1996}, \cite{mawatari_critical_1997}, \cite{mikitik_analytical_2013}. However, these solutions are only valid under specific operating conditions such as AC transport current or externally applied magnetic field. Therefore, they can not be directly used to estimate losses in most superconducting machines and equipment.  For these reasons, a finite element model is typically used to analyze the electromagnetic behavior and estimate hysteretic losses in the HTS tapes. 

There are two main formulations of Maxwell's equations that are commonly used to model superconductors by using finite elements method (FEM). The first one is based on the magnetic field strength (\textbf{H}) and has already been used to study numerous applications \cite{shen_review_2020}, \cite{shen_overview_2020}. The second one was introduced in \cite{zhang_efficient_2016} and is based on the current vector potential (\textbf{T}) and magnetic vector potential (\textbf{A}). This T-A formulation is mostly used to analyze superconducting layers by applying a thin strip approximation. This approach reduces degrees of freedom and computation time. Therefore, it has been used to study the cross-section of magnets \cite{berrospe-juarez_estimation_2018} and electrical machines with hundreds and thousands of tapes \cite{benkel_t-formulation_2020}, \cite{huang_fully_2020}, \cite{vargas-llanos_t-formulation_2020}. 

Most of the FEM-based models used to study superconducting devices are 2D. They usually represent the cross-section of an infinite long or axisymmetric arrangements. Therefore, the end effects are not considered. Moreover, the current 3D models require high computation times to simulate single coils in commercially available computers  \cite{zhang_efficient_2016}, \cite{zermeno_3d_2014}.

In this work, we use the 3D T-A homogenization and show its effectiveness in simulating 3D objects like HTS racetrack coils.
The 3-D homogenized T-A formulation was first proposed by Huang et al. \cite{huang_effective_2019}, who used it to show its superiority in calculating AC losses of HTS racetrack coils with respect to 2D axisymmetric models. Here, we first  describe the full derivation of the governing equations, with particular emphasis on their implementation in COMSOL Multiphysics (section~\ref{Section_T_A_Formulation}). Then, we validate the 3D T-A formulation model with measurements and axisymmetric solutions (section \ref{Section_circular_coil}). The approach is used to study a racetrack coil that cannot be fully analyzed with 2D models in section \ref{Section_racetrack_coil}. These simulations are compared with the estimation of losses presented in \cite{zermeno_3d_2014} by showing that the 3D T-A homogenization can  reproduce the same results with a significant reduction of computation time. Finally, the main conclusions of this work are summarized in section \ref{Conclu}.

\section{T-A Formulation and Homogenization}\label{Section_T_A_Formulation}

In this section we first present a small review of the T-A formulation and homogenization to set the basis that allows us to extend these concepts to 3D. Then, the 3D T-A homogenization is explained and its implementation in circular and rectangular geometries/sections is presented.

\subsection{T-A formulation}
The first applications of the T-A formulation to study superconducting devices by using FEM were introduced in \cite{zhang_efficient_2016}, \cite{liang_finite_2017}. These works presented the T-A formulation as an efficient approach to model HTS tapes with a high aspect ratio.  

This formulation of Maxwell's equations couples the current vector potential \textbf{T} and magnetic vector potential \textbf{A}, which are defined by the current density \textbf{J} and magnetic flux density \textbf{B}:

\begin{equation} 
\mathbf{J}=\nabla \times \mathbf{T} \label{T} 
\end{equation}

\begin{equation} 
\mathbf{B}=\nabla \times \mathbf{A}. \label{A}
\end{equation}

\begin{figure}
\centerline{\includegraphics[width=0.5\textwidth]{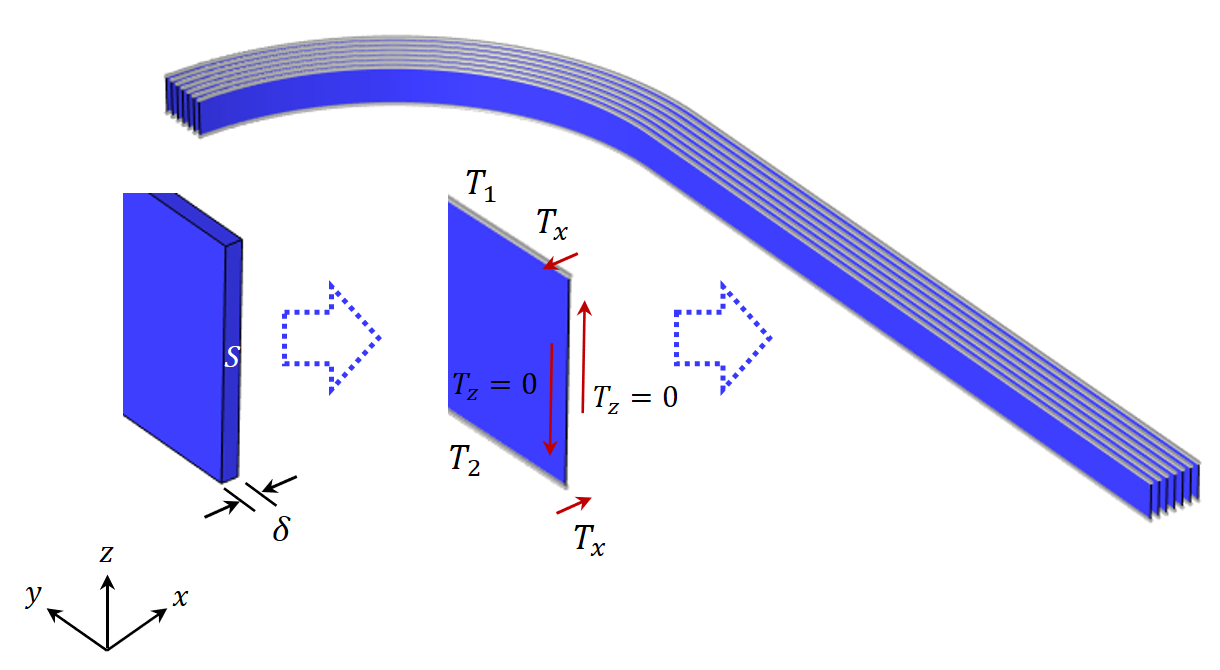} }
\caption{T-A formulation in 3D. \textbf{T} is computed only in the superconducting domain (depicted in blue) while \textbf{A} is computed everywhere. The tape's thickness is collapse into a superconducting layer and the current is enforced by giving proper values of $T$ at the edges of the tape \cite{berrospe-juarez_real-time_2019}.}
\label{T-A_formulation_3D}
\end{figure}

The magnetic vector potential is computed in all the domains under study by using Maxwell-Ampere's law ($\mu$ is the relative permeability of the material):

\begin{equation} \label{eq_Amp_A_formulation}
\nabla \times (\frac{1}{\mu} \nabla \times \mathbf{A}) = \mathbf{J}. \end{equation}

The current vector potential is calculated only in the superconducting domain by using Maxwell-Faraday's law:

\begin{equation} \label{eq_T_Faraday}
\nabla \times (\rho _{\rm HTS} \nabla \times \mathbf{T}) = -\frac{\partial \mathbf{B}}{\partial t}.
\end{equation}

The resistivity of the superconducting material is typically modeled with a power-law \textbf{E-J} relation \cite{rhyner_magnetic_1993}:

\begin{equation} \label{eq_HTS_resis}
\rho _{\rm HTS}=\frac{E_{\rm c}}{J_{\rm c}(\mathbf{B})}  \Bigg| \frac{\mathbf{J}}{J_{\rm c}(\mathbf{B})} \Bigg| ^{n-1}.
\end{equation}

An approximation is done by considering that the superconducting layer in the tapes under study (for instance rare-earth barium copper oxide/REBCO tapes) has a very large width-to-thickness ratio. Therefore, we can collapse the thickness of the tape ($\delta$) as shown in figure~\ref{T-A_formulation_3D}. As a consequence, the current will only be able to flow in a superconducting sheet and  \textbf{T} will always be perpendicular to this sheet. For this reason, the current vector potential can be expressed as $T \cdot \textbf{n}$ ($\textbf{n}$ is a unitary vector perpendicular to the superconducting layer)
\cite{zhang_efficient_2016}.


We can obtain the required boundary conditions at the edges of the tape/layer by integrating the current:

\begin{equation} \label{eq_enfo_I1}
I=\iint_S \mathbf{J} \,{\rm d}S=\iint_S \nabla \times \mathbf{T} \,{\rm d}S=\oint_{\partial S} \mathbf{T} \,{\rm d}l.
\end{equation}



As it can be seen in figure~\ref{T-A_formulation_3D}, equation (\ref{eq_enfo_I1}) can be reduced into: 

\begin{equation} \label{eq_enfo_I2}
I=(T_1-T_2)\delta,
\end{equation}
which allows enforcing the current in coils by giving proper values of $T$ at the edges of each superconducting tape. 


\subsection{Homogenization in 3D}

\begin{figure}
\centerline{\includegraphics[width=0.5\textwidth]{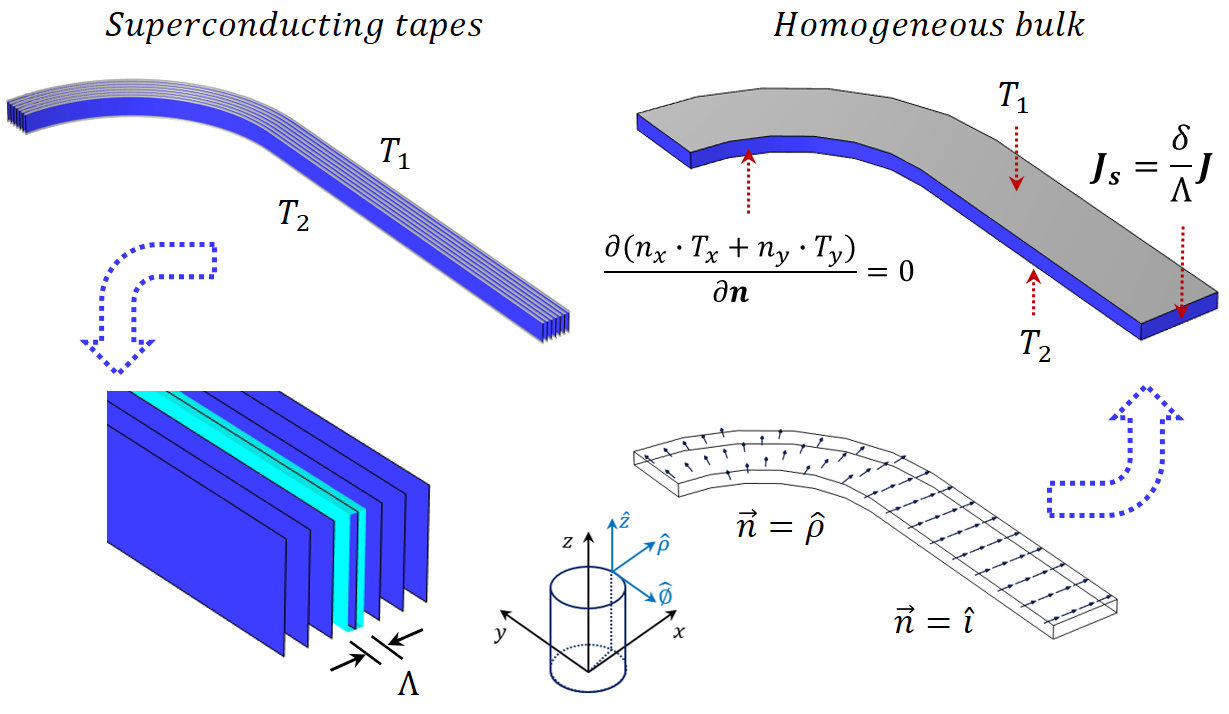} }
\caption{T-A homogenization in 3D. The superconducting sheets are replaced by a homogeneous block. The scale current density $\textbf{J}_s$ is introduced as a source term and boundary conditions $T_1$ and $T_2$ are applied to the upper and lower boundaries.}
\label{T-A_homogenization_3D}
\end{figure}

The homogenization technique assumes that the superconducting tapes that are wound in coils can be represented by an anisotropic bulk that is able to reproduce the overall electromagnetic behavior of the coil \cite{zermeno_calculation_2013}. This allows reducing degrees of freedom and computation time. The technique was implemented in 2D by using the T-A formulation in \cite{berrospe-juarez_real-time_2019}, \cite{2020arXiv200602033B}. Therefore, we follow a similar approach to extend the homogenization into the 3D analysis of superconducting coils.

\begin{figure*}[hbt!]
  \centering
  \includegraphics[width=0.7\textwidth]{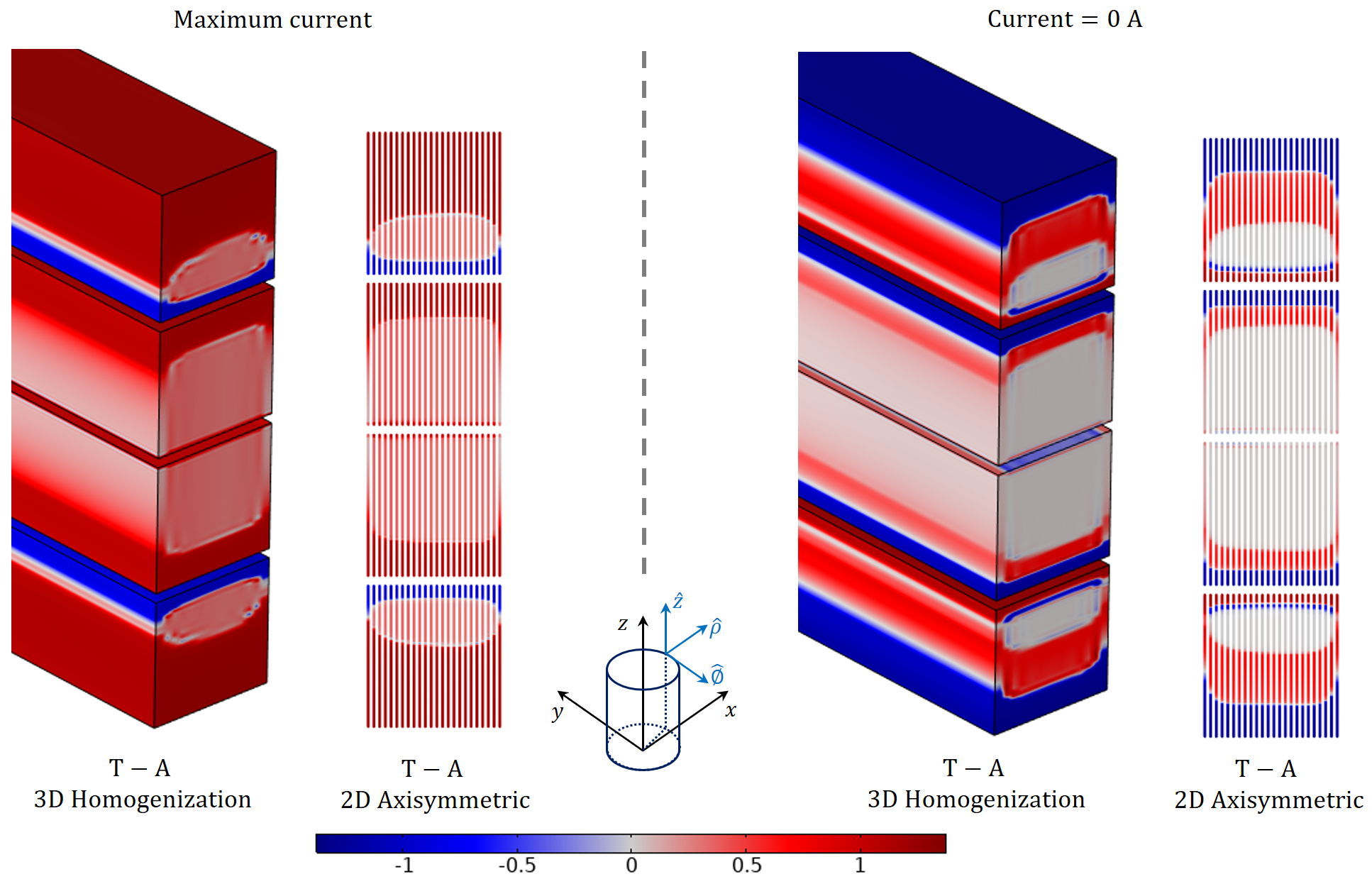}
  \caption{Normalized current density behavior in four circular coils stacked one on top of the other for transport current (55 A peak) without externally applied magnetic field. A comparison between the T-A 3D homogenization and T-A 2D axisymmetric solution is done when the current reaches the maximum and zero values.}
  \label{Circular_current}
\end{figure*}

The procedure is summarized in figure~\ref{T-A_homogenization_3D}. We start from the arrangement of tapes modeled in the T-A 3D formulation as sheets and define a cell unit around them (with the same height of the superconducting tape and thickness $\Lambda$). The main characteristics of the tape will be impressed in this cell unit to transform the stack of tapes into a block. The scaled current in the homogenized bulk is defined for the \textbf{A} calculation as:

\begin{equation} \label{eq_Js}
\textbf{J}_{s} = \frac{\delta}{\Lambda} \textbf{J}. \end{equation}

In principle, the homogenization makes use of Maxwell's equations in the same way as the 3D T-A formulation. Therefore, we keep the thin strip approximation of the tape that allows reducing the current vector potential into a scalar quantity. This means that even if the tapes are replaced by a homogeneous bulk, the current can only flow in the plane parallel to the original superconducting sheets. For example,  in the zoom presented in the bottom left corner of figure~\ref{T-A_homogenization_3D}, the current can only have a $J_y$ and $J_z$ component and the current vector potential will only have a $T_x$ component. The new homogenized block can be seen as a highly compressed group of superconducting tapes.
To represent a general geometry, we assume $\textbf{n}=\begin{bmatrix} n_x \\ n_y \\ n_z\end{bmatrix}$ and express equation (\ref{T}) as: 

\begin{equation} 
\begin{bmatrix} J_x \\ J_y \\ J_z\end{bmatrix}
=
\begin{bmatrix} 
\frac{\partial(T \centerdot n_z)}{\partial y} - \frac{\partial(T \centerdot n_y)}{\partial z} \\
\frac{\partial(T \centerdot n_x)}{\partial z} - \frac{\partial(T \centerdot n_z)}{\partial x} \\
\frac{\partial(T \centerdot n_y)}{\partial x} - \frac{\partial(T \centerdot n_x)}{\partial y}
\end{bmatrix}.
 \label{T2} 
\end{equation}

The magnetic vector potential is calculated in all the domains and we solve the current vector potential only in the superconducting domain by using Maxwell-Faraday's law:

\begin{equation} 
\begin{bmatrix} 
\frac{\partial(E_z)}{\partial y} - \frac{\partial(E_y)}{\partial z} \\
\frac{\partial(E_x)}{\partial z} - \frac{\partial(E_z)}{\partial x} \\
\frac{\partial(E_y)}{\partial x} - \frac{\partial(E_x)}{\partial y} 
\end{bmatrix} 
\centerdot\textbf{n}
+
\begin{bmatrix} 
\frac{\partial(B_x)}{\partial t}\\
\frac{\partial(B_y)}{\partial t}\\
\frac{\partial(B_z)}{\partial t}
\end{bmatrix}
\centerdot\textbf{n}
=0,
\label{eq_T_Faraday2} 
\end{equation}
where $\textbf{n}$ can be easily determined in the T-A formulation because it is the vector perpendicular to the superconducting sheet. Therefore, it is usually defined by default in commercial software like COMSOL Multiphysics. However, once the stack of tapes is replaced, it is not easy to define $\textbf{n}$ inside the bulk. We have no longer a reference surface from which we can define the normal vector. A similar issue can be found in the 2D T-A homogenization when complex (rotated or curved) cross-sections are modeled in 2D. For this reason, we followed the geometrical path of the tape before homogenization to define the vector. If the tape is parallel to the $y-z$ plane then:

\begin{equation} 
\textbf{n}=\hat i=\begin{bmatrix} 1 \\ 0 \\ 0\end{bmatrix}.
 \label{n_straight} 
\end{equation}

If the tape is wound in a circular shape with the center in the origin, then $\textbf{n}$ will be parallel to the radial vector in cylindrical coordinates $\hat \rho$:   
 
\begin{equation} 
\textbf{n}=\hat \rho=\begin{bmatrix} \frac{x}{\sqrt{x^2+y^2}} \\ \frac{y}{\sqrt{x^2+y^2}} \\ 0\end{bmatrix}.
 \label{n_circular} 
\end{equation} 

The normal vector can also be defined by domains to represent more complex geometries as the one shown in the right corner of figure~\ref{T-A_homogenization_3D}. In this case, $\textbf{n}=\hat i$ in the straight section and $\textbf{n}=\hat \rho$ in the circular one.

Finally, we have to establish the boundary conditions to solve our problem. The homogenized bulk represents a densely packed group of HTS sheets. Each one of these sheets should transport the same current as its original counterpart \cite{berrospe-juarez_real-time_2019}. Therefore, we apply Dirichlet boundary conditions to the upper and lower gray boundaries as expressed in equation (\ref{eq_enfo_I2}) and apply Neumann boundary conditions to the internal and external boundaries:

\begin{equation} 
\frac{\partial (n_x \centerdot T_x + n_y \centerdot T_y)}{\partial \textbf{n}}=0.
 \label{eq_Neumann_BC} 
\end{equation}

\section{Modelling of Circular Coils}\label{Section_circular_coil}

The first analyzed case with the 3D T-A homogenization is the circular coils presented in \cite{pardo_electromagnetic_2015}. This geometry can be studied with an axisymmetric model. Therefore, we can compare not only losses but also the normalized current density distribution and overall electromagnetic behavior between the 3D T-A homogenization and the 2D T-A formulation. In addition to this comparison, we can validate the model with the measurements shown in \cite{pardo_electromagnetic_2015}.

The coils under study have an internal radius of \SI{30}{\milli\meter}, external radius  \SI{33.9}{\milli\meter} and 24 turns. They are wound with a \SI{4}{\milli\meter} REBCO tape that has a \SI{1}{\micro\meter} superconducting layer. The critical current of the tape at \SI{77}{\kelvin} and self-field is \SI{128}{\ampere}. More information about the coils and measurements can be found in \cite{pardo_electromagnetic_2015}. In these simulations, we assumed a critical electric field $E_c=\SI{1e-4}{\volt\per\meter}$ and $n=25$.

Since the resistivity of the superconductor is several orders of magnitude lower than the resistivity of other materials in the tape, we model the HTS tape in 2D by considering only the superconducting layer surrounded by a homogeneous medium with a resistivity of \SI{1}{\ohm \meter} and vacuum permeability. This approximation is valid for low-frequency applications like the transport losses of the circular coils under analysis\cite{musso_analysis_2021}.

The critical current density dependence on the magnetic field magnitude and direction is modeled as:

\begin{equation} 
J_{\rm c}(B_{\parallel},B_{\bot})=\frac{J_{\rm c0}}{1+\frac{\sqrt{k^{2}B_{\parallel}^{2}+B_{\bot}^{2}}}{B_{\rm c0}}}.\label{eq_JcB}
\end{equation}

In this equation, $B_{\parallel}$ and $B_{\bot}$ are the parallel and perpendicular components (to the flat face of the tape) of the magnetic flux density. The parallel component is calculated as the norm of the two components of the magnetic flux density parallel to the tape ($B_{\parallel}=\sqrt{B_z^2+B_\phi^2}$) and the perpendicular component is the one aligned with the normal vector ($B_{\bot}=B_\rho$). The parameters $J_{\rm c0}$, $k$ and $B_{\rm c0}$ are equal to \SI{3.2e10}{\ampere\per\square\meter}, \SI{0.25}{} and \SI{150}{\milli\tesla}.  

We analyze first the behavior of one single coil. Then, we simulated groups of two, three and four coils stacked one on top of the other with AC transport current ($f=\SI{36}{\hertz}$) and without an external magnetic field. Figure~\ref{Circular_current} shows the behavior of the normalized current density ($J/J_{\rm c}$) for the four coils arrangement and peak transport current of \SI{55}{\ampere}. We can see in this picture that the current distribution inside the coils is very similar between the 3D T-A homogenization and the 2D T-A axisymmetric model when the current is maximum and zero. Similar behavior was observed for other instants in time and different groups of coils. 

\begin{figure}
\centerline{\includegraphics[width=0.5\textwidth]{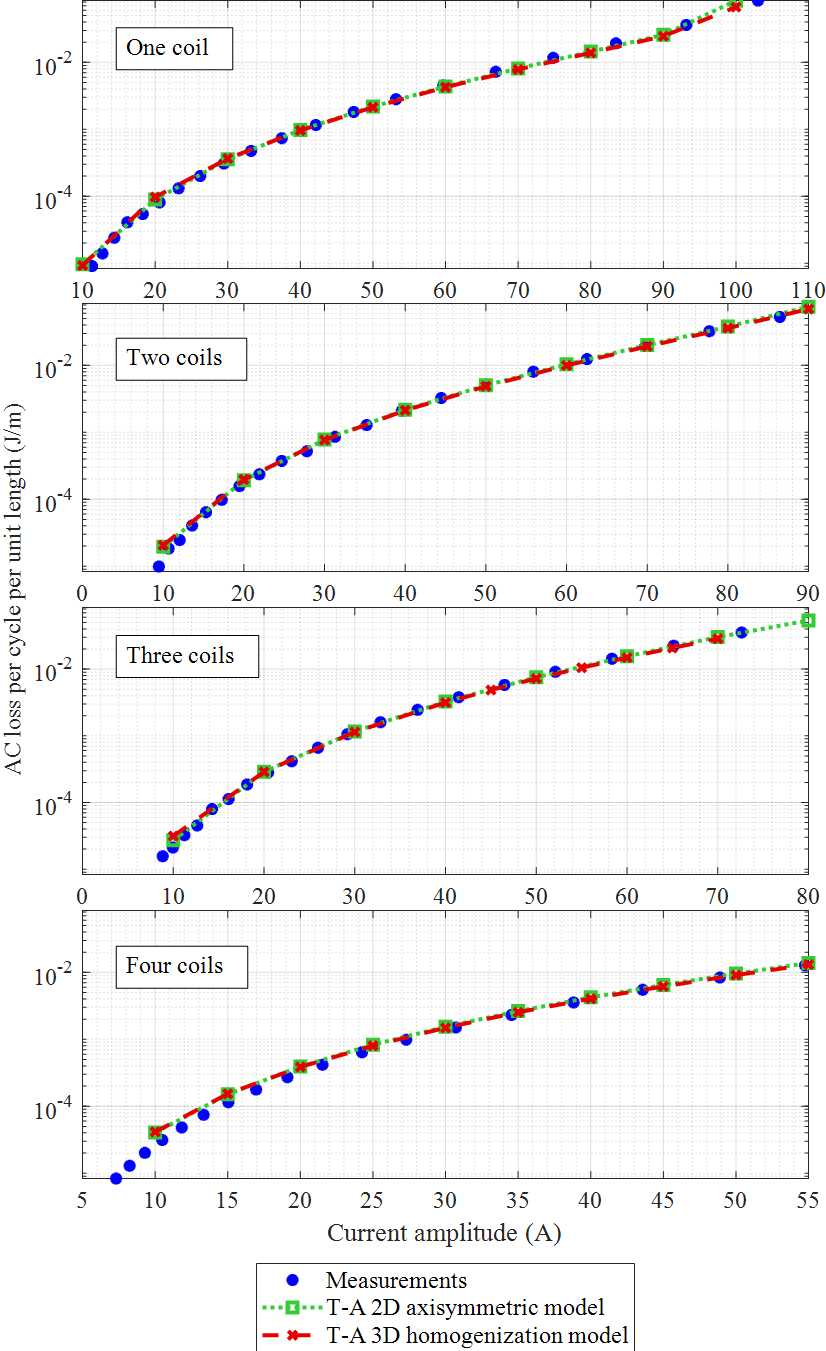} }
\caption{Estimation of losses due to AC transport current in the circular coils by using the T-A 2D axisymmetric, the T-A 3D homogenization and measurements. The calculation was done for a single coil and groups of two, three and four coils stacked one on top of the other.}
\label{Circular_coils}
\end{figure}

Figure~\ref{Circular_coils} shows a comparison of the losses measured in \cite{pardo_electromagnetic_2015} and estimated with the 3D T-A homogenization and 2D T-A axisymmetric models for all the coil's arrangements. As it can be seen, the 3D T-A homogenization is in good agreement with the 2D axisymmetric model and measurements for all the studied current ranges and coils arrangements.

\section{Modelling of a Racetrack Coil}\label{Section_racetrack_coil}

Since the modeling technique was already validated in the previous section with 2D axisymmetric models, we can proceed to analyze more complicated geometries that can not be investigated in 2D.

\begin{figure}[hbt!]
\centerline{\includegraphics[width=0.4\textwidth]{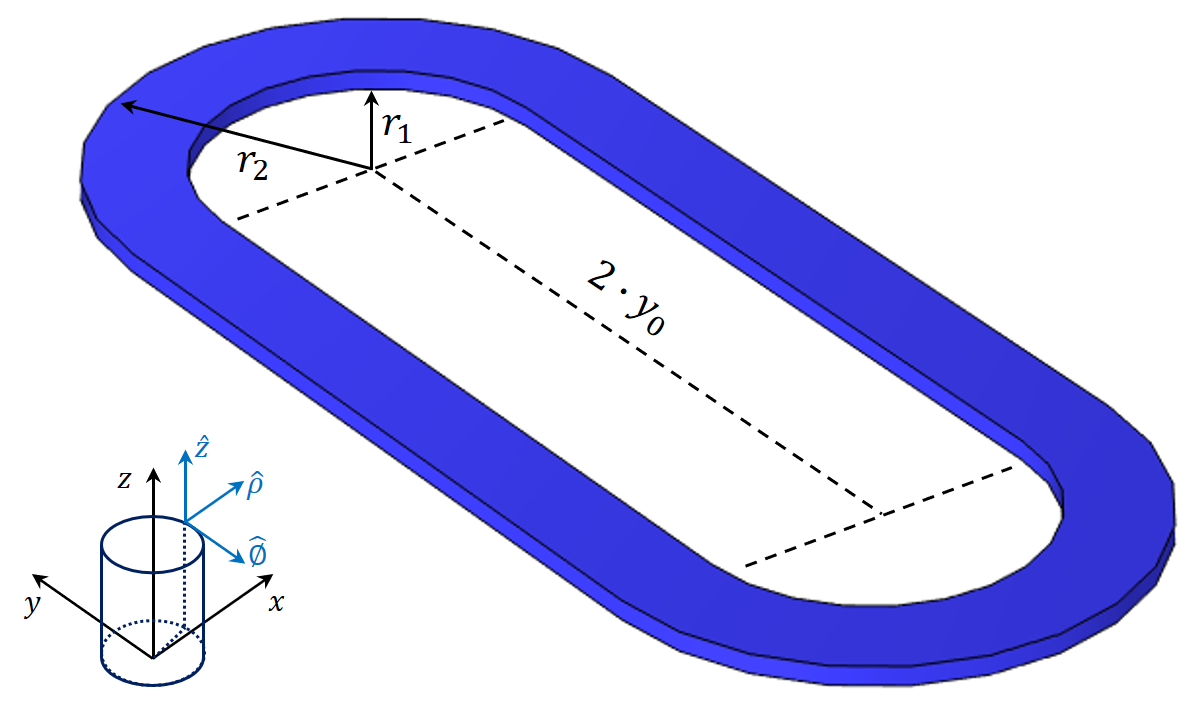} }
\caption{Geometry of the racetrack coil under study.}
\label{Racetrack_geom}
\end{figure}

In this section, we study the racetrack coil presented in \cite{zermeno_3d_2014}. This kind of coils is typically used in superconducting electrical machines for wind turbine applications \cite{abrahamsen_large_2012}. These coils have two straight and two round parts as shown in figure~\ref{Racetrack_geom}. For this reason, the normal vector (\textbf{n}) must be defined by domain following the original path of the superconducting tape as it was mentioned in section \ref{Section_T_A_Formulation}. 

The dimensions of the coil under study are summarized in table~\ref{table_dimensions_racetrack}. This coil is wound with a \SI{4}{\milli\meter} tape with a critical current of \SI{160}{\ampere} at self-field. We assume in this section the same parameters presented in \cite{zermeno_3d_2014} to model the superconducting tape. Similarly to what was done in \cite{zermeno_3d_2014}, we modeled only one-eighth of the coil by taking advantage of the symmetries. This allows a fair comparison between the T-A 3D and H 3D homogenization techniques.    

\begin{table}[hbt!]
\centering
\caption{Dimensions of the racetrack coil.}
\label{table_dimensions_racetrack}
\begin{tabular}{l c} 
\hline			
Internal radius ($r_1$)                     &	\SI{35}{\milli\meter}	\\
External radius ($r_2$)                     &	\SI{55}{\milli\meter}	\\
Straight part length ($2 \centerdot y_0$)    &	\SI{150}{\milli\meter}	\\
Thickness                                    &	\SI{4}{\milli\meter}	\\
Turns                                        &	50	\\
\hline
\end{tabular}
\end{table}

\begin{figure*}[hbt!]
  \centering
  \includegraphics[width=0.8\textwidth]{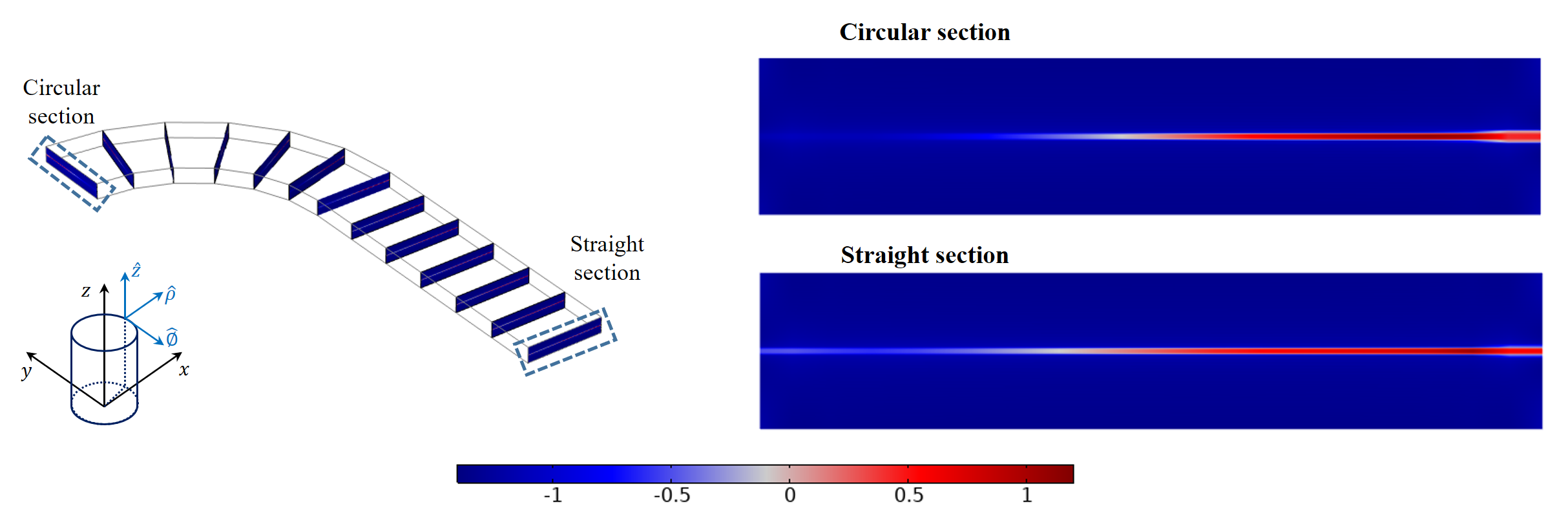}
  \caption{The behavior of the normalized current density in one-fourth of the racetrack coil for AC transport current ($f=\SI{50}{\hertz}$ and $I_{\rm peak}=\SI{120}{\ampere}$) when the current is closed to the negative maximum. A Zoom in the middle of the circular section and the straight section of the racetrack coil is presented on the right.}
  \label{Racetrack_J_Jc}
\end{figure*}

\begin{figure}
\centerline{\includegraphics[width=0.5\textwidth]{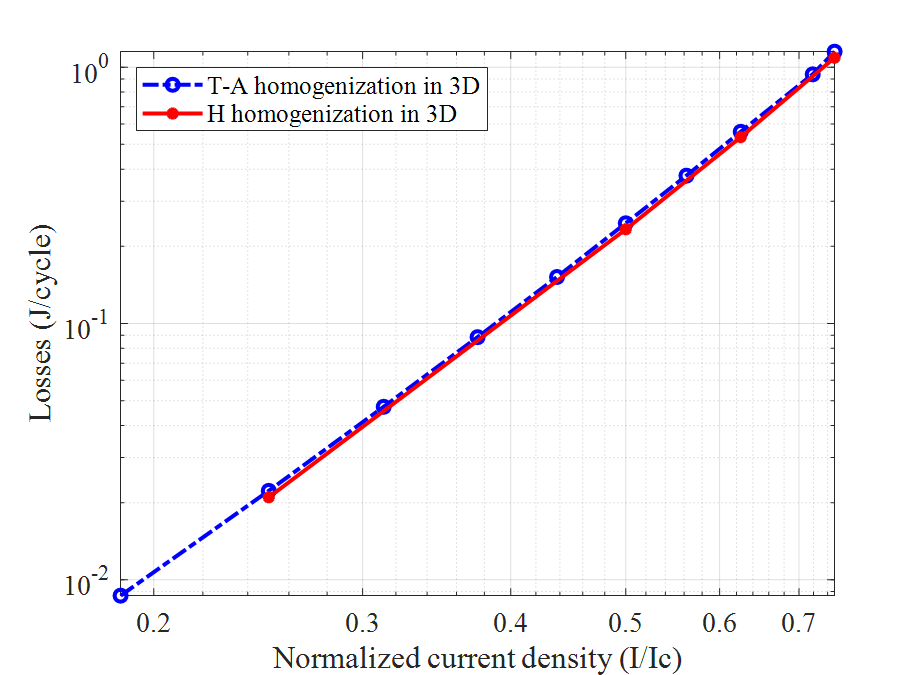} }
\caption{Estimation of losses due to AC transport current ($f=\SI{50}{\hertz}$) as a function of the current amplitude by using the T-A 3D homogenization and the H 3D homogenization. The current is normalized with respect to the critical current of the tape at \SI{77}{\kelvin}.}
\label{Racetrack_coil_losses}
\end{figure}

The critical current density dependence on the magnetic field magnitude and direction was modeled by using the parallel ($B_{\parallel}$) and perpendicular ($B_{\bot}$) components of the magnetic flux density \cite{zermeno_3d_2014}. For this reason, these components were also calculated by domain. In the circular section, they were computed  as $B_{\parallel}=\sqrt{B_z^2+B_\phi^2}$ and  $B_{\bot}=B_\rho$, by assuming that the circular region is centered at the origin. In the straight section, they were computed  as $B_{\parallel}=\sqrt{B_z^2+B_y^2}$ and  $B_{\bot}=B_x$, by considering that the straight region is parallel to the $y$ axis.

The behaviour of the normalized current density in the cross-sections of the racetrack coil for AC transport current ($f=\SI{50}{\hertz}$ and $I_{\rm peak}=\SI{120}{\ampere}$) is presented in figure~\ref{Racetrack_J_Jc}. At first sight, it seems that the current density distribution is very similar inside the coil. However, if we make a zoom in the middle of the circular and the straight section of the coil when we are close to the maximum current (positive or negative), we can notice that the innermost part of the circular section fills first with the critical current. This behavior can be related to a higher magnetic flux density in this internal area of the coil, in comparison with the straight region. Therefore, the analysis of the behavior of this section can be decisive in the operation of the coil. 

Figure~\ref{Racetrack_coil_losses} shows the losses due to AC transport current (\SI{50}{\hertz}) estimated with the 3D T-A homogenization model. As it can be seen, the results are in good agreement with the 3D H homogenization calculation for all the studied current values. 

The main advantage of this new 3D modeling technique is a significant reduction in computation time. Table~\ref{Computation_time} presents a comparison between the computation times of the 3D homogenized T-A and H models for a similar number of degrees of freedom: 178484 and 176138 for T-A and H-formulation, respectively. These calculations were made with a \SI{3.60}{\giga\hertz} Intel Core i7-4960X computer with 12 logical processors and \SI{64}{\giga\byte} of RAM. As can be observed, the T-A 3D homogenization is faster and achieves a reduction in computation time between \SI{80.43}{\percent} and \SI{81.64}{\percent}. 
It should be remarked that the mesh of the T-A formulation was coarser than that of the H formulation and that a comparison of the computational efficiencies of the two formulations needs further work.
However, one advantage of the the T-A 3D homogenization is that it is relatively easy to implement: in particular, it does not require 2D integral constraints or high resistivity zones to prevent current sharing between subdomains of different groups of tapes. This restriction, which is quite tricky to implement in the 3D H homogenization~\cite{zermeno_3d_2014}, is included in the essence of the T-A homogenization by applying the thin strip approximation which constraints the current to flow in the plane parallel to the tape and allows reducing the current vector potential into a scalar quantity.  


\begin{table}[hbt!]
\centering
\caption{Comparison of the computation time between the T-A and H homogenization in 3D for the simulations of the racetrack coil. The current is normalized with respect to the critical current of the tape at \SI{77}{\kelvin} and self-field.}
\label{Computation_time}
\begin{tabular}{ P{1.5cm} P{1.5cm} P{1.5cm} P{1.5cm}} 
\hline
\hline
Normalized current (I/Ic) &  Time H formulation & Time T-A formulation & Percentage reduction \\
\hline
\SI{0.25}{} & \SI{17.38}{\hour} & \SI{3.19}{\hour} & 81.64 \\
\SI{0.5}{} & \SI{29.74}{\hour} & \SI{5.66}{\hour} & 80.98 \\
\SI{0.625}{} & \SI{35.16}{\hour} & \SI{6.76}{\hour} & 80.77 \\
\SI{0.75}{} & \SI{41.18}{\hour} & \SI{8.06}{\hour} & 80.43 \\
\hline
\hline
\end{tabular}
\end{table}

\section{Conclusion}\label{Conclu}

A 3D T-A homogenization modeling technique was used to study the behavior of superconducting coils. The modeling approach was applied to the circular coils described in \cite{pardo_electromagnetic_2015}. This allowed validating the homogenization technique assumptions with 2D axisymmetric solutions, where a similar electromagnetic behavior was observed. Furthermore, the estimation of losses was corroborated with measurements which verifies the accuracy of the model.  

Finally, the approach was applied to the study of a racetrack coil. This kind of coil cannot be fully analyzed with a 2D model. The study showed that the 3D T-A homogenization results are in good agreement with the 3D H homogenization and the approach has the potential to achieve a significant reduction in computation time. This opens the way to modeling and analyzing more complex 3D superconducting configurations in commercially available computers within a reasonable computation time.





\section*{Acknowledgment}
The authors thank Dr. Enric Pardo for sharing measured data of the transport losses in the circular coils.

\bibliographystyle{IEEEtran}
\bibliography{My_Library_23032021.bib}

\end{document}